\documentclass{aa}  
\usepackage{graphicx}
\usepackage{txfonts}
\newcommand{\excs}{\extracolsep{\fill}} 
\begin{document}
   \title{Characterization of the long-period companions of the exoplanet host stars: HD\,196885, HD\,1237 and HD\,27442\thanks{Based on ESO observing programs 075.C-0825(A), 275.B-5057A and 077.C-0444(A)}}

   \subtitle{VLT/NACO and SINFONI near-infrared, follow-up imaging and spectroscopy}

\author{
        G. Chauvin\inst{1, 2}
	\and A.-M. Lagrange\inst{2}
	\and S. Udry \inst{3}  
	\and M. Mayor \inst{3}}

 \offprints{Ga\"el Chauvin \email{gchauvin@eso.org}}
\institute{ $^{1}$European Southern Observatory, Casilla 19001, Santiago 19, Chile\\ 
$^{2}$Laboratoire d'Astrophysique, Observatoire de Grenoble, UJF, CNRS; BP 53,
  F-38041 GRENOBLE Cdex 9 (France)\\
$^{3}$Observatoire de Gen\`eve, 51 Ch. des Maillettes, 1290 Sauverny,
Switzerland\\ }

   \date{Received December 29, 2006; accepted March 5, 2007}

 
  \abstract 
{} 
{We present the results of near-infrared, follow-up imaging and
spectroscopic observations at VLT, aimed at characterizing the
long-period companions of the exoplanet host stars HD\,196885,
HD\,1237 and HD\,27442. The three companions were previously
discovered in the course of our CFHT and VLT coronographic imaging
survey dedicated to the search for faint companions of exoplanet host
stars.}
{We used the NACO near-infrared adaptive optics instrument to obtain
astrometric follow-up observations of HD\,196885\,A and B. The
long-slit spectroscopic mode of NACO and the integral field
spectrograph SINFONI were used to carry out a low-resolution spectral
characterization of the three companions HD\,196885\,B, HD\,1237\,B
and HD\,27442\,B between 1.4 and 2.5~$\mu$m.}
{We can now confirm that the companion HD\,196885\,B is comoving with
its primary exoplanet host star, as previously shown for HD\,1237\,B
and HD\,27442\,B. We find that both companions HD\,196885\,B and
HD\,1237\,B are low-mass stars of spectral type M$1\pm1$V and
M$4\pm1$V respectively. HD\,196885\,AB is one of the
closer ($\sim23$~AU) resolved binaries known to host an
exoplanet. This system is then ideal for carrying out a combined radial velocity and
astrometric investigation of the possible impact of the binary
companion on the planetary system formation and evolution. Finally, we
confirm via spectroscopy that HD\,27442\,B is a white dwarf
companion, the third one to be discovered orbiting an exoplanet host star,
following HD\,147513 and Gliese~86. The detection of the broad Br$\gamma$
line of hydrogen indicates a white dwarf atmosphere dominated by
hydrogen.}
{}

   \maketitle
%

\section{Introduction}

The radial velocity (RV) technique is without contest nowadays the
most successful method for detecting and characterizing the properties
of exo-planetary systems. Since the discovery of 51 Peg (Mayor \&
Queloz 1995), more than 200 exo-planets have been identified featuring
a broad range of physical (mass) and orbital (P, $e$) characteristics
(Udry et al. 2003). Major progress has been made in improving
detection performance and data analysis and has enabled us to explore
the mass regime down to neptunian and telluric masses (Santos et
al. 2004; McArthur et al. 2004). Hitherto, most surveys have been
focused on solar-type stars because these stars show more and thinner
lines than their more massive counterparts and less activity than
their less massive ones, which ensures in both cases comparatively
higher RV precision. Only recently have been planet-search programs
devoted to late-type stars with the use of high precision spectrograph
and repeated measurements (Delfosse et al. 1998; Endl et al. 2003;
Wright et al. 2004; Bonfils et al. 2005) as well as to early-type
stars with the development of new methods for radial velocity
measurements (Chelli 2000; Galland et al. 2005).

Despite the success of this technique, the time span explored limits
the study to the close ($\le4-5$~AU) circumstellar environment. To
understand the way exo-planetary systems form and evolve, it is therefore
clearly worthwhile using complementary techniques such as pulsar
timing, micro-lensing, photometric transit or direct imaging to
fill out our knowledge. For solar analogs, the current deep-imaging
capability is limited to the detection of massive brown dwarf (BD)
companions. Typical separations larger than $50-100$~mas
(i.e. $\geq 3-5$~AU from a star at 50~pc) can be explored in the stellar
regime and $0.5-1~\!''$ (i.e. $\geq 30-50$~AU from a star at 50~pc) in the
substellar regime. The recent discovery of a T7 dwarf companion at
480~AU from the exoplanet host star HD\,3651 (Mugrauer et al. 2006)
illustrates the scope of deep, near-infrared imaging for resolving
ultra-cool companions at large separations. Recent efforts have been
also devoted to systematic search for stellar companions to nearby
stars with and without planets, aimed at studying the impact of
stellar duplicity on planet formation and evolution (Eggenberger et
al. 2007).

Since 2003, we have conducted a deep-coronographic imaging survey of
26 exoplanet host stars, using PUEO-KIR at CFHT, and NACO at VLT
(Chauvin et al. 2006). Three probable companions were detected around
the stars HD\,196885, HD\,1237 and HD\,27442 (see
Tables~\ref{tab:tab1} and \ref{tab:tab2}). Follow-up observations were
obtained for HD\,1237\,B and HD\,27442\,B, confirming their
companionship (Chauvin et al. 2006; Raghavan et al. 2006). Based on
their photometry, HD\,196885\.B and HD\,1237\,B are likely to be
low-mass stars and HD\,27442\,B is probably the third white dwarf
companion known to date orbiting an exoplanet host star. In this
paper, we report new imaging and spectroscopic observations of these
three companions, using NACO and SINFONI at VLT. In section~2, the
instrument set-up and the data analysis are described. In section~3,
we discuss the astrometric and spectroscopic results that confirm
and refine their previous suggested nature.

\begin{table}[t]
\caption{Characteristics of the observed exoplanet host stars HD\,196885\,A, HD\,1237\,A and HD\,27442\,A.}
\label{tab:tab1}
\centering
\begin{tabular*}{\columnwidth}{@{\excs}lllllll}     
\hline\hline
Name         &  SpT         & d         &  Age$^a$  &   $\rm{M}_{2}\rm{sin}i$$^b$   &  $P\,^b$  &   $e\,^b$     \\        
             &              & (pc)      &  (Gyr)    &   $(M_{\rm{Jup}})$        &  (days)  &            \\

         \\ 
\hline
HD\,196885\,A   & F8IV         & 33.0      & 0.5       &   1.84                 & 386.0  & 0.30 \\
HD\,1237\,A     & G6V          & 17.6      & 0.8       &   1.94                 & 311.29 & 0.24\\
HD\,27442\,A    & K2IV         & 18.2      & 10        &   1.28                 & 423.84 & 0.07 \\
\hline
\end{tabular*}
\begin{list}{}{}
\item[\scriptsize{($^a$) AGE REFERENCES:}] \scriptsize{Naef et al. 2001; Randich et al. 1999; Lambert et al. 2004}
\item[\scriptsize{($^b$) RADIAL VELOCITY REFERENCES:}]\scriptsize{Naef et al. 2001; Butler et al. 2001; http://exoplanets.org/esp/hd196885/hd196885.shtml}
\end{list}
\end{table}


\section{Observations and data reduction}

\subsection{NACO Imaging of HD\,196885\,AB} 

The NACO adaptive optics (AO) instrument is installed at the Nasmyth B
focus of VLT/UT4 and provides diffraction-limited images in the
near-infrared ($1-5~\mu$m) range. On 26 August 2006, we used NACO to
carry out a second-epoch observation of HD\,196885\,A and B and confirm that
both systems were co-moving. The atmospheric conditions were stable
although not optimal (airmass of 1.3, seeing of $\omega=0.9~\!''$ and
correlation time of $\tau_0=3.0$~ms). A set of five jittered images was
obtained using the K$_{s}$ filter and the S27 camera CONICA,
leading to a total exposure time of $\sim2$~min on source. To
calibrate the plate scale and the detector orientation, we observed at
each epoch the astrometric field of $\theta$\,Ori\,1\,C (McCaughrean \&
Stauffer 1994). After cosmetic reductions using \textit{eclipse}
(Devillard 1997), we used the deconvolution algorithm of V\'eran \&
Rigaut (1998) to obtain the position of HD\,196885\,B relative to A at
each epoch. Single stars of similar brightness observed the same night
were used for point spread function (PSF) estimation. The results are reported in
Table~\ref{tab:astrohd196885} and the relative positions plotted in Fig.~\ref{fig:astro}.

\begin{table}[t]
\caption{Characteristics of the long-period companions to the exoplanet host stars HD\,196885\,A, HD\,1237\,A and HD\,27442\,A. Date, separation and reference of the first observation are reported. The contrast values in $K_{s}$ come from Chauvin et al. (2006).}
\label{tab:tab2}
\centering
\begin{tabular*}{\columnwidth}{@{\excs}lllllll}     
\hline\hline
 Name        &UT Date     &   $\Delta$   &  $\Delta$    &    $\Delta K_{s}$  & Ref.  \\
             &            & (arcsec)        & (AU)         & (mag)         &           \\
\hline
HD\,196885\,B  & 01/08/2005   & $0.714$    &  25          & $3.1$        & (1)   \\
HD\,1237\,B  & 03/06/2003   & $3.857$    &    68        &  $5.0$       & (1)   \\
HD\,27442\,B  & 1930   & $\sim13.7$    &   240         & $10.7$        & (2)   \\
\hline
\end{tabular*}
\begin{list}{}{}
\item[\scriptsize{($1$):}] \scriptsize{Chauvin et al. 2006}
\item[\scriptsize{($2$):}]\scriptsize{Worley \& Douglass 1997}
\end{list}
\end{table}

\subsection{NACO spectroscopy of HD\,196885\,B and HD\,1237\,B} 

NACO also provides AO-assisted, long-slit, near-infrared
spectroscopy. We used the low-resolution ($R_{\lambda}=550$) grism
with the SHK ($1.40-2.50$~$\mu$m) filter and the 86~mas slit to
characterize HD\,196885\,B on August 27 2006 and HD\,1237\,B
on July 7 2005. A standard ABBA nodding sequence in and out the slit
was conducted to properly remove the background contribution. The
telluric standard stars HIP\,104320 (B3V) and HIP\,007873 (B2V) were
observed for HD\,196885\,B and HD\,1237\,B respectively. After
substracting the sky and dividing by a flat field using
\textit{eclipse}, the science and reference spectra
were extracted and calibrated in wavelength using the
REDSPEC\footnote{http://www2.keck.hawaii.edu/inst/nirspec/redspec/}
software.

The system HD\,1237 was observed at high airmass (1.8) and under poor
atmospheric conditions (seeing $\omega=1.0~\!''$ and correlation time
of $\tau_0=2.5$~ms) resulting in a poor and unstable AO
correction. These unstable conditions have a critical effect on the
continuum slope of the observed spectra when long-slit spectroscopy is
coupled to AO (see Goto et al. 2002; Chauvin et al. 2005). To
determine the spectral type of HD\,1237\,B, we then decided to
use a limited spectral range between 2.3 and 2.4~$\mu$m where the
obvious temperature-sensitive features of the CO molecular bands are
present. The continuum slope of each spectrum was fitted between 2.20
and 2.45~$\mu$m and divided. The result is shown in
Fig.~\ref{fig:spechd1237}. In the case of HD\,196885\,A and B, the
atmospheric conditions were stable (airmass of 1.2, seeing of
$\omega=0.8~\!''$ and correlation time of $\tau_0=3.5$~ms). As the
flux ratio between both components was relatively small, we placed
both objects into the 86~mas slit. The reduced spectra in H and K-band
are shown in Fig.~\ref{fig:spechd196885}.

\begin{table}[t]
\caption{Offset positions of HD\,196885\,B relative to A measured with NACO at VLT.}
\label{tab:astrohd196885}
\centering
\begin{tabular*}{\columnwidth}{@{\excs}llllllll}     
\hline\hline
 UT Date     &   $\Delta$   &  P.A.    &    Plate scale  &  Detector Orien. \\
             & (mas)        & (deg)    & (mas)              & (deg)          \\
\hline
01/08/2005   & $714\pm3$    & $67.5\pm0.3$  & $27.01\pm0.05$     & $-0.05\pm0.20$    \\
26/08/2006   & $713\pm3$    & $65.7\pm0.3$  & $27.01\pm0.05$     & $-0.12\pm0.19$    \\
\hline
\end{tabular*}
\end{table}

\subsection{SINFONI spectroscopy of HD\,27442\,B} 

SINFONI is a near-infrared integral field spectrograph fed by an AO
module, currently installed at the Cassegrain focus of the VLT/UT4. To
characterize the companion HD\,27442\,B, we used the medium resolution
($R_{\lambda}=1500$) grating with the H+K ($1.45-2.45~\mu$m) filter,
coupled to the 25~mas objective (providing a field of view of
$0.8~\!''\times0.8~\!''$). In comparison with NACO, the great
advantage of SINFONI is that the instrument is relatively insensitive
to differential chromatical effects as the whole PSF is dispersed. The
observations were obtained on October 18 2005. The telluric standard
star HIP\,39483 (B3V) was successively observed. The SINFONI pipeline
and the QFitsView\footnote{http://www.mpe.mpg.de/~ott/QFitsView/}
software were used to properly reduce the observations (sky
subtraction, flat-fielding, wavelength calibration and extraction) and
the reduced spectrum of HD\,27442\,B is shown in
Fig.~\ref{fig:spechd196885}.


\section{Results}

\subsection{HD\,196885\,B, a close M1V dwarf companion}

\begin{figure}[t]
\centering
\includegraphics[width = \columnwidth]{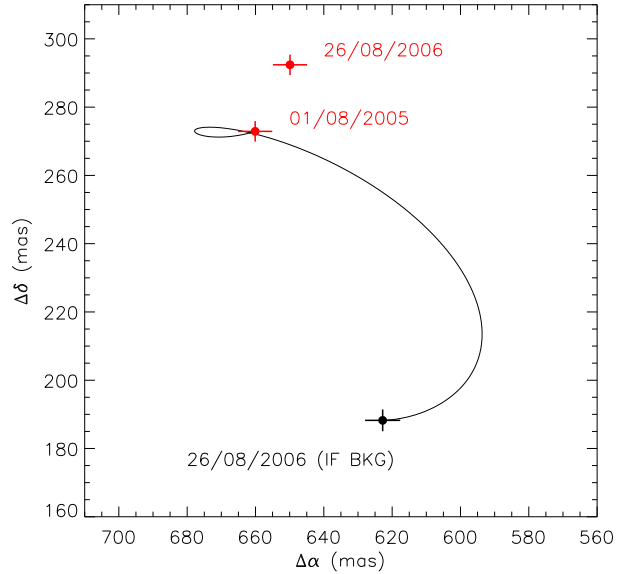}
\caption{VLT/NACO measurements with uncertainties in the offset
positions of HD\,196885\,B relative to A, obtained on August 1st 2005
and August 26 2006. The expected variation of offset positions, if B
is a background stationary object, is shown (\textit{solid line}),
based on a distance of $33\pm1$~pc, a proper motion of
$(\mu_{\alpha},\mu_{\delta})=(47.5\pm0.9, 83.1\pm0.5)$~mas/yr for A
and the initial offset position of B from A. The corresponding
expected offset positions of a background object on August 26 2006 is
also given with total uncertainties, which are the quadratic sum
of individual uncertainties (distance, proper motion of A and initial
position of B from A).}
\label{fig:astro}
\end{figure}

The exoplanet host star HD\,196885\,A was first observed on August 1st
2005 and a relatively bright companion candidate (here after
HD\,196885\,B) was resolved at $0.7~\!''$. By re-imaging this system
at a second epoch, our goal was to show that HD\,196885\,B shares a
common proper motion with A. Fig.~\ref{fig:astro} displays, in a
($\Delta\alpha$, $\Delta\delta$) diagram, the offset positions of
HD\,196885\,B from A, observed with NACO on August 1st 2005 and August
26 2006. The expected evolution of the relative A-B positions, under
the assumption that B is a stationary background object, is indicated
for August 26 2006. Based on the relative positions of B from A at
both epochs (see Table~\ref{tab:astrohd196885}), the $33\pm1$~pc
distance of A and its proper motion of
$(\mu_{\alpha},\mu_{\delta})=(47.5\pm0.9, 83.1\pm0.5)$~mas/yr, we find
that the $\chi^2$ probability that HD\,196885\,B is a background
stationary object is less than 1e-9. HD\,196885\,B is then likely to
be comoving with A.

The JHK photometry of HD\,196885\,B is compatible with that expected
for an M0V to M1V dwarf (see Chauvin et al. 2006; Leggett et
al. 1996). The NACO H-band and K-band spectra of HD\,196885\,B feature
neutral atomic lines for MgI (1.50 and 1.71~$\mu$m), AlI
(1.68~$\mu$m), NaI (2.20 and 2.33~$\mu$m) and CaI (2.26~$\mu$m) as
well as weak molecular lines for CO and H$_2$O, typical of early-M
dwarfs. Using the template spectra of Pickles et al. (1998) and
Leggett et al. (2001), the best fit to the H- and K-band spectra of
HD\,196885\,B is obtained using the M1V dwarf LHS386
spectrum. Therefore, photometry and spectroscopy confirm a spectral
type M$1\pm1$V for HD\,196885\,B. Using Baraffe et al. (1998) model
predictions for an age of 0.5~Gyr, we derive a mass of
$0.5-0.6$~M$_{\odot}$, which gives a period of several decades. It is
therefore not surprising that our two epoch measurements already
resolve the orbital motion of B relative to A (see
Fig.~\ref{fig:astro}).

With a physical projected separation of 23~AU, this system is among
the closer binaries known to host a planetary system. Like Gliese\,86
(Lagrange et al. 2006), HD\,41004 (Zucker et al. 2004) and $\gamma$\,Cet
(Neuh\"auser et al. 2007),
HD\,196885 is an ideal system for carrying out combined astrometric
and RV observations to constrain the binary dynamic properties and
the possible impact of a close binary companion on planet
formation and evolution. 

\begin{figure*}[t]
\centering
\includegraphics[width = 9.2cm]{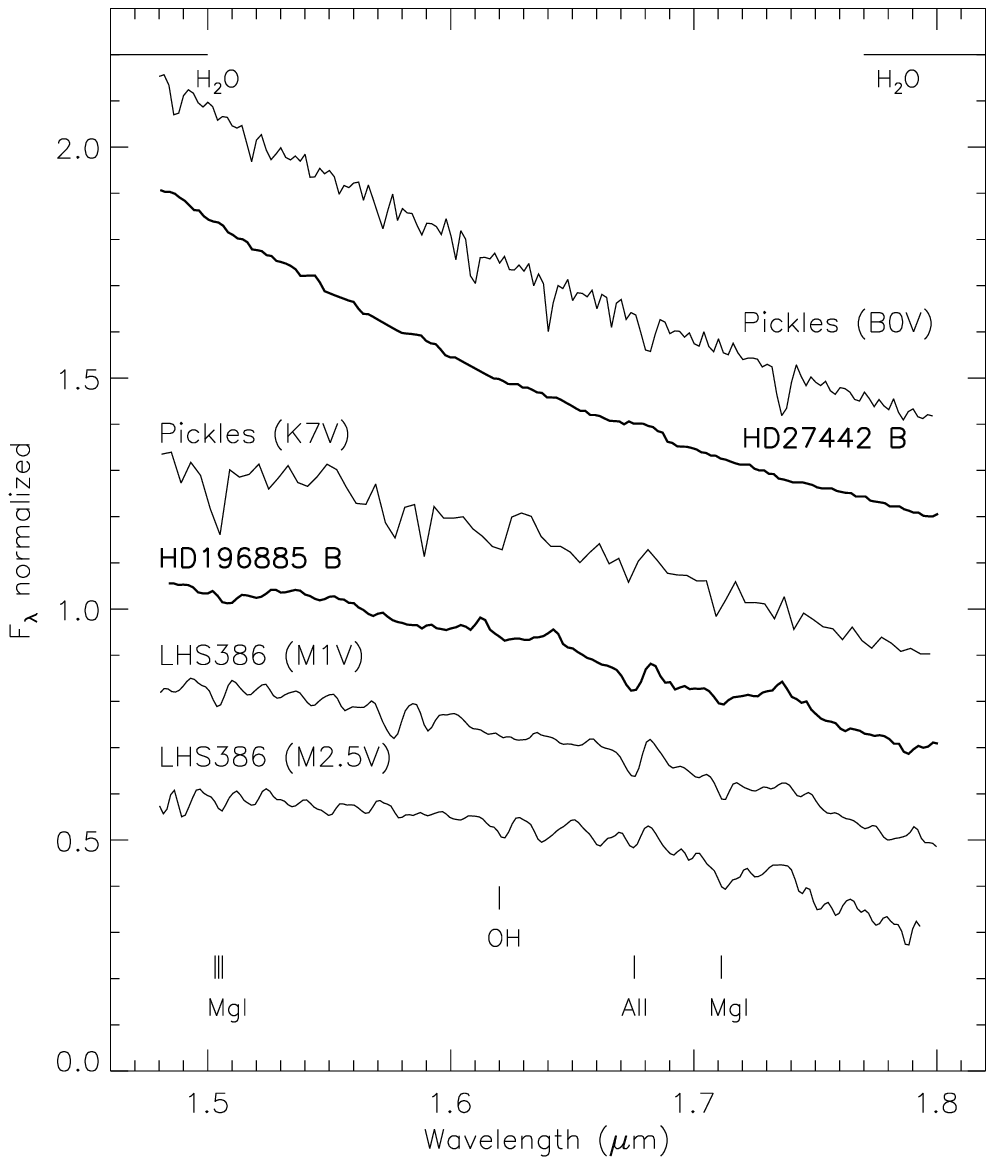}\includegraphics[width = 9.2cm]{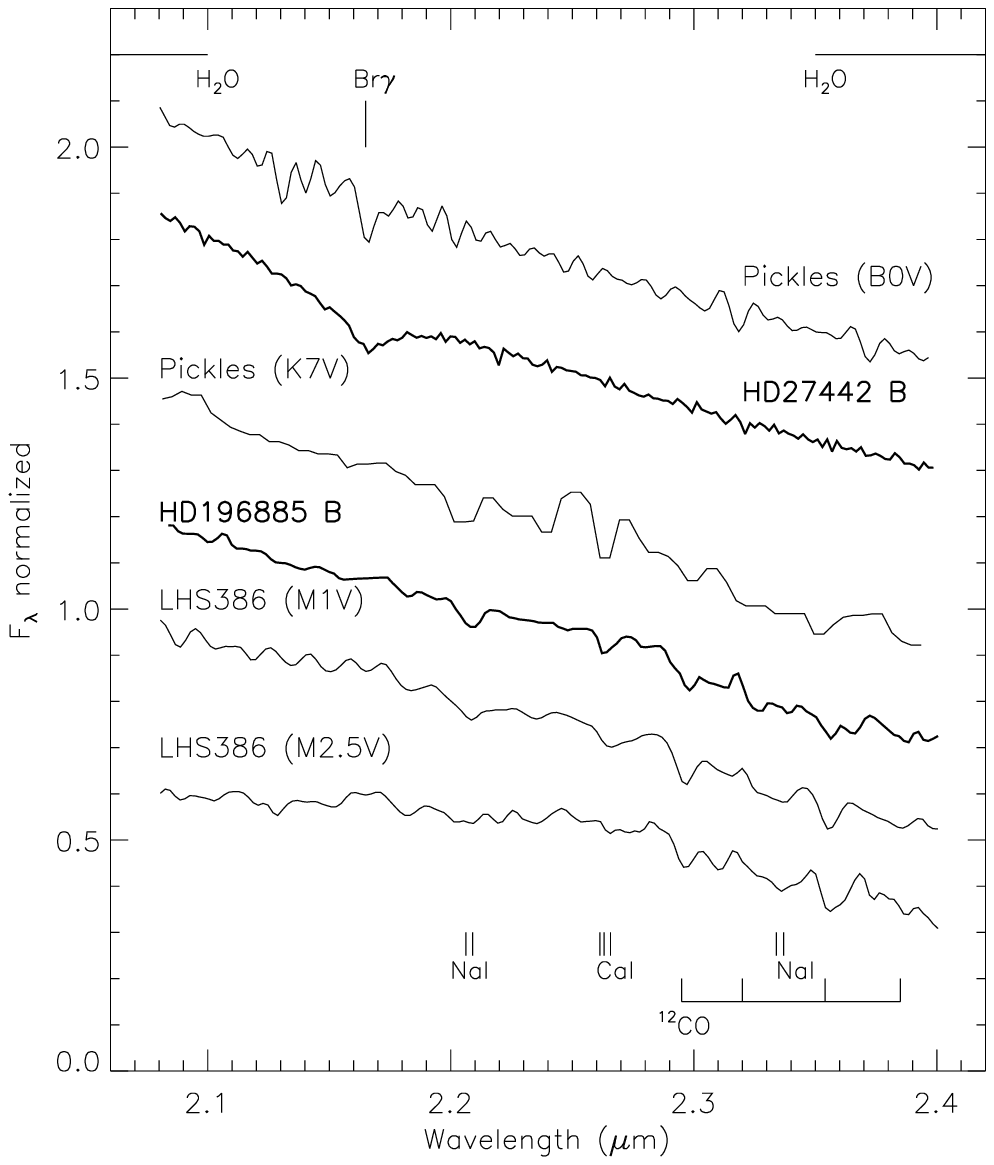}
\caption{VLT/NACO and SINFONI spectroscopy for HD\,27442\,B and
HD\,196885\,B over the spectral range (1.4-1.9~$\mu$m, \textit{left})
and (2.0-2.5~$\mu$m, \textit{right}). Template spectra of Pickles et
al. (1998) and Leggett et al. (2001) are shown for direct
comparison.}
\label{fig:spechd196885}
\end{figure*}

\subsection{HD\,1237\,B, an M4V dwarf}

HD\,1237\,B is located at a projected angular separation of 3.8~$\!''$
(i.e a projected physical separation of 68~AU) from the exoplanet host
star HD\,1237\,A. Previous NACO follow-up observations showed that
HD\,1237\,B is comoving with A and revealed the orbital motion of B
relative to A (Chauvin et al. 2006). The near-infrared
photometry of HD\,1237\,B is compatible with that expected for an M4V
to M6V field dwarfs according to the evolutionary model predictions of
Baraffe et al. (1998).

\begin{figure}[t]
\centering
\includegraphics[width = \columnwidth]{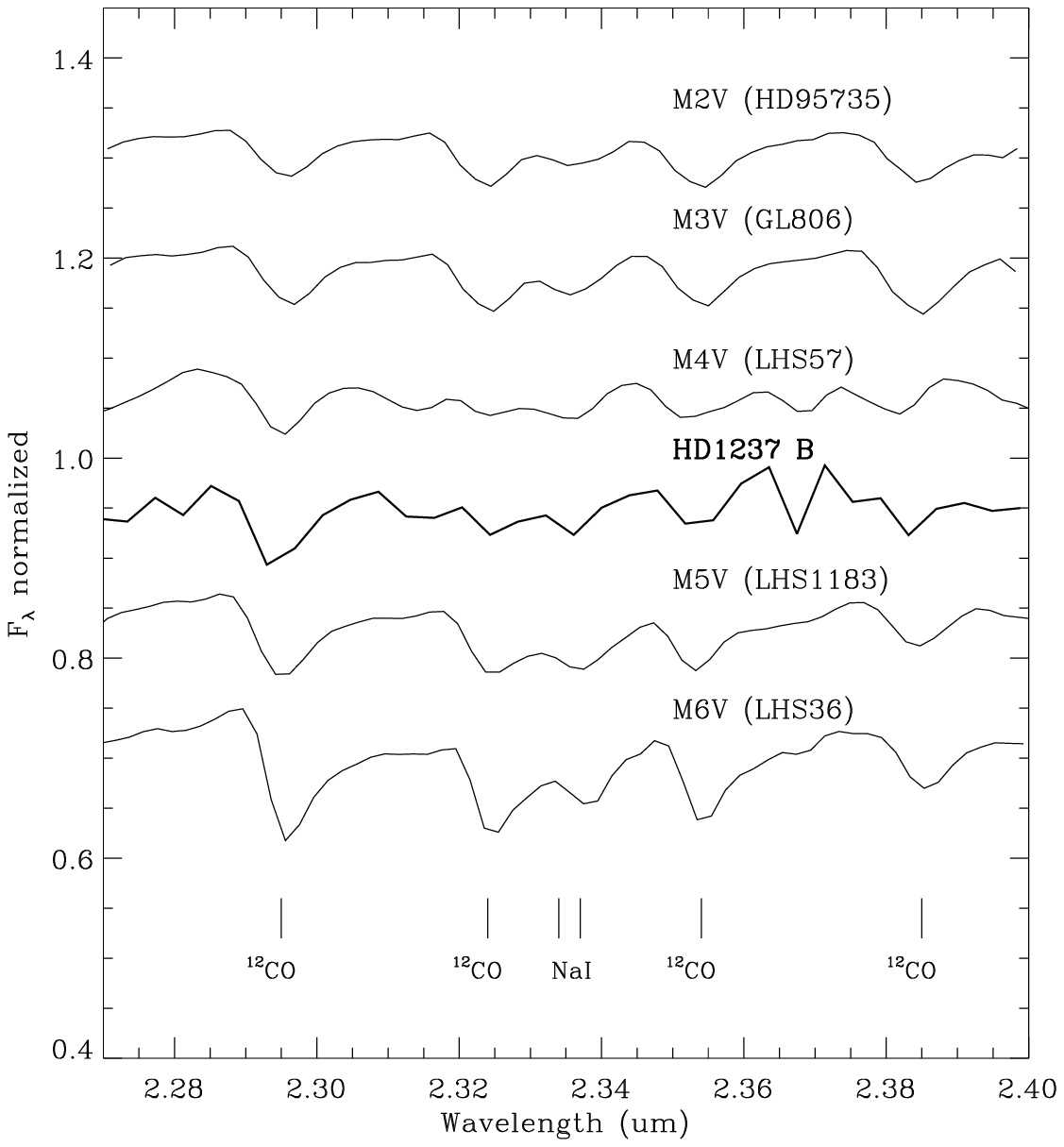}
\caption{VLT/NACO near-infrared spectrum of HD\,1237\,B obtained with
the low resolution ($R_{\lambda}=550$) grism, the SHK
($1.40-2.50$~$\mu$m) filter and the 86~mas slit. The limited spectral
range between 2.27 and 2.40~$\mu$m is shown. For spectral type
determination, the obvious temperature sensitive features of the CO
molecular bands were compared to the template spectra of M dwarfs
(Leggett et al. 2001, Cushing et al. 2005), smoothed to the same
spectral resolution}. This comparison confirms a spectral type
M$4\pm1$V for HD\,1237\,B.
\label{fig:spechd1237}
\end{figure}

Based on the near-infrared spectroscopy, the HD\,1237\,B
spectrum can be compared to the template spectra of M dwarfs (Leggett
et al. 2001; Cushing et al. 2005; see
Fig.~\ref{fig:spechd1237}). Using the CO index that measure the
strength of the $^{12}$CO 2-0 band head at 2.29~$\mu$m (McLean et
al. 2003), we find a value of 0.945 which excludes spectral types
later than M6V according to Cushing et al. (2005). Using a minimum
$\chi^2$ adjustment to find the best spectra matching the CO bands of
our HD\,1237\,B spectrum, we derive a spectral type M$4\pm1$V for
HD\,1237\,B. Additional neutral atomic lines typical of M dwarfs are
also present such as MgI (1.50~$\mu$m) NaI (2.20 and 2.33~$\mu$m) and CaI
(2.26~$\mu$m). This result is also in agreement with the spectral type
estimation independently obtained by Mugrauer et al. (2007) using ISAAC
at the VLT.

For a system age of 0.8~Gyr, the predictions of the Baraffe et
al. (1998) model give a mass of $\sim0.13$~M$_{\odot}$. The
maximum radial velocity drift that B should produce on A is about
$4$~m.s$^{-1}$.yr$^{-1}$ (assuming a circular orbit). The
chromospheric activity of HD\,1237\,A induces a RV residual of 18
m\,s$^{-1}$ (Naef et al. 2001), which excludes a short-term study
combining RV with astrometric measurements to dynamically characterize
this system.

\subsection{HD\,27442\,B, a DA white dwarf}

On November 17 2003, the faint companion HD\,27442\,B was detected at
$\sim13~\!''$ (i.e 240~AU in projected physical distance) from the
subgiant exoplanet host star HD\,27442\,A. Follow-up observations
confirmed that HD\,27442\,A and B are comoving (Raghavan et
al. 2006, Chauvin et al. 2006). The combined use of NACO JK
photometry and V-band observations from the Washington Visual Double
Star Catalog (WDS, Worley \& Douglass 1997) revealed that HD\,27442\,B
is not a main sequence star or a brown dwarf, but more probably a
white dwarf companion. The SINFONI near-infrared spectrum for
HD\,27442\,B spectral range confirms unambiguously this assumption
(see Fig.~\ref{fig:spechd196885}). The H and K-band spectra do not
present any H$_2$O or CO molecular bands typical of cool stellar or
substellar atmospheres. The continuum slope follows the Rayleigh-Jeans
domain typical of warm, early-type star atmospheres. The lack
of helium lines and the detection of a broad Br$\gamma$
(2.166~$\mu$m) hydrogen feature indicates a white-dwarf atmosphere
dominated by hydrogen (DA). The broad Br$\gamma$ line profile results
from pressure and Doppler broadening and is a well-known
characteristic of white dwarfs' dense atmospheres. While writing this
paper, a similar analysis was obtained by Mugrauer et al. (2007) using
ISAAC at the VLT who independently confirm the hydrogen-rich white
dwarf status of HD\,27442\,B.

Holberg et al. (2002) found that 70\% of the nearby
($d\le20$~pc) white dwarfs are hydrogen-rich and 25\% are part of
multiple systems, which does not mean that the HD\,27442 (K2IV\,+\,DA)
binary system is anything odd. The main particularity is that the
primary is a subgiant exoplanet host star and is in that sense a more
evolved system than HD\,147513 (G5V\,+\,DA, Porto de Mello \& da Silva
1997, Mayor et al. 2004) and Gliese 86 AB (K1V\,+\,WD; Queloz et
al. 2000, Els et al. 2001, Jahreiss 2001), which were the first two
systems of this kind to be discovered.

Using visible and near-infrared photometry on HD\,27442\,B compared to
predictions from the evolutionary model of Bergeron et al. (2001) for
white dwarfs with hydrogen-rich atmospheres, the mass of HD\,27442\,B
ranges from 0.3 to 1.2~M$_{\odot}$ and the effective temperature from
9000 to 17000~K. The next step would be to carry out a detailed
photometric and spectroscopic analysis in optical and near-infrared
wavelengths to determine the effective temperature and the surface
gravity of HD\,27442\,B. Such an analysis can be done based on the
comparison of hydrogen Balmer lines or combined optical BVRI and
infrared JHK photometry with model predictions for pure hydrogen
atmospheres (Bergeron et al. 1992; 1997). As the trigonometric
parallax of the system is known thanks to HD\,27442\,A, the radius of
HD\,27442\,B and hence its mass could be determined through the
theoretical mass-radius relation for white dwarfs.

%
\section{Conclusions}

We present the follow-up imaging and spectroscopic characterization of
three long-period companions to the exoplanet host stars HD\,196885,
HD\,1237 and HD\,27442. The three objects were discovered during a
previous deep-imaging survey carried out at CFHT and at VLT. Our new
observations confirm their companionship unamiguously as well as their
nature, which had previously been inferred from their photometry. Whereas
HD\,196885 and HD\,1237 are two stellar companions of
spectral type M$1\pm1$V and M$4\pm1$V respectively, HD\,27442 is the second
confirmed white dwarf companion of an exoplanet host star. The
detection of the broad Br$\gamma$ hydrogen line indicates a white
dwarf atmosphere dominated by hydrogen.

HD\,196885\,AB is one of the closer resolved binaries known to host an
exoplanet. The presence of this long-period companion may have played
a key-role in the formation and the evolution of the inner planetary
system, but how? The two main mechanisms of planetary formation, core
accretion and disk instability, do not lead to the same predictions
while the impact of a binary companion, depending on the separation
and the mass ratio, is still debated. A complete dynamic
characterization of nearby, tight binaries with planets and a
dedicated imaging survey to study the multiplicity among stars with
and without planets are clearly required to throw new light
on the mechanisms of planetary formation and evolution.

\begin{acknowledgements}

We would like to thank the staff of ESO-VLT for their support at the
telescope and Sandy Leggett and Michael Cushing for providing us with their
library of near-infrared template spectra of M dwarfs. This publication
has made use of the SIMBAD and VizieR database operated at CDS, Strasbourg, France. Finally, we acknowledge partial
financial support from the {\sl Programmes Nationaux de Plan\'etologie
et de Physique Stellaire} (PNP \& PNPS) and the {\sl Agence Nationale
de la Recherche}, in France.

\end{acknowledgements}

\end{document}